\documentclass[%
 reprint,
superscriptaddress,
 amsmath,
 amssymb,
 aps,
 pra,
longbibliography
]{revtex4-2}

\usepackage{graphicx}
\usepackage{dcolumn}
\usepackage{bm}
\usepackage{xcolor}

\usepackage[hidelinks,colorlinks=true,linkcolor=blue,citecolor=blue,urlcolor=blue]{hyperref}

\usepackage{bbm}

\usepackage[]{algorithm2e}

\usepackage{lipsum}

\makeatletter
\newcommand*{\balancecolsandclearpage}{%
  \close@column@grid
  \clearpage
  \twocolumngrid
}
\makeatother

\begin{document}

\preprint{APS/123-QED}

\title{Stochastic cloaking: concealing a region from diffusive particles}

\author{Connor Roberts}
\email{connor.roberts16@imperial.ac.uk}
\affiliation{Department of Mathematics, Imperial College London, SW7 2AZ, United Kingdom}

\author{Ziluo Zhang}
\affiliation{Department of Mathematics, Imperial College London, SW7 2AZ, United Kingdom}
\affiliation{Wenzhou Institute, University of Chinese Academy of Sciences, Wenzhou, Zhejiang 325001, China}

\author{Helder Rojas}
\affiliation{Departamento de Matemáticas Fundamentales, Universidad Nacional de Educación a Distancia, Calle de Juan del Rosal 10, 28040 Madrid, Spain}

\author{Stefano Bo}
\affiliation{Department of Physics, King's College
London, WC2R 2LS, United Kingdom}

\author{Carlos Escudero}
\affiliation{Departamento de Matemáticas Fundamentales, Universidad Nacional de Educación a Distancia, Calle de Juan del Rosal 10, 28040 Madrid, Spain}

\author{S\'{e}bastien Guenneau}
\email{s.guenneau@imperial.ac.uk}
\affiliation{Department of Physics, Imperial College London, SW7 2AZ, United Kingdom}

\author{Gunnar Pruessner}
\email{g.pruessner@imperial.ac.uk}
\affiliation{Department of Mathematics, Imperial College London, SW7 2AZ, United Kingdom}

\date{\today}

\begin{abstract}
We introduce ``stochastic cloaking,'' where a region of space is concealed from an ensemble of diffusing particles whose individual trajectories are governed by a stochastic (Langevin) equation. Our simulations reveal how different interpretations of the Langevin equation affect the cloaking performance of an annular single-layer invisibility cloak of smoothly varying diffusivity in two dimensions. Near-perfect cloaking is achieved under the It\^{o} convention, indicated by the cloak preventing particles from accessing an inner core without disturbing the particle density outside the cloak. The cloak's performance can be further improved by regularising its singular behaviour. We believe our demonstration of stochastic cloaking is a significant milestone, comparable to earlier developments that extended cloaking from optics and acoustics to thermodynamics.
\end{abstract}

\maketitle

\paragraph*{Introduction.}
The advent of ``metamaterials'' has led to the proposal of various phenomena that were once considered the realm of science fiction \cite{shelby2001experimental, smith2000composite, greenleaf2009cloaking, leonhardt2006optical}. Notably, by capitalising on the form-invariance of Maxwell's equations under a coordinate transformation \cite{ward1996refraction}, Pendry suggested that tuning the magnetic permeability and electric permittivity of a metamaterial could be used to deviate electromagnetic radiation around an object in such a way that the radiation is undisturbed from its original trajectory \cite{pendry2006controlling}. The metamaterial acts as an invisibility cloak in this case, since any external observer would be unable to detect the presence of the concealed object and, crucially, the metamaterial itself. This application of ``transformation optics'' has inspired research into cloaking for other forms of radiation. 

A major milestone is thermal cloaking, based on ``transformation thermodynamics'' \cite{guenneau2012transformation, schittny2013experiments, raza2016transformation}, where a metamaterial of heterogeneous conductivity protects a region from heat flux. Thermal cloaking demonstrated that transformation-based cloaking applies not only to elliptic differential equations---describing optics and acoustics \cite{cummer2007one, milton2006cloaking}---but also to parabolic differential equations. Yet, both are examples of \textit{deterministic} differential equations. 

Till now, no equivalent demonstration has been made for systems governed by \textit{stochastic} differential equations, where behaviour can be predicted only on a \textit{statistical} (rather than \textit{precise}) basis. Such a demonstration would be another significant milestone for cloaking, not least because of the inherent difficulties introduced by stochasticity. The major difficulty arises from the governing stochastic equation---the Langevin equation---being uniquely determined only after choosing a particular convention under which to integrate it \cite{van1981ito}. This is often called the It\^{o} vs.\ Stratonovich dilemma \cite{van1981ito, van1992stochastic, escudero2023versus, yuan2012beyond, mannella2012ito}, named after the first two conventions to be proposed \cite{ito1944109, stratonovich1966new}. Though both are over 60 years old, their physical interpretation remains of great interest today \cite{pacheco2024langevin}.

Switching conventions typically involves a simple transformation of the Langevin equation \cite{lau2007state}. However, this requires a degree of regularity that turns out to be absent in our model, similarly to the one-dimensional cases of Refs.~\cite{correales2019ito,escudero2020kinetic,escudero2023fluctuation,escudero2023versus}. This negates the ability to readily switch conventions and thus makes the original choice even more critical. Additionally, we consider a multi-dimensional model, which poses further mathematical challenges as well as important implications for noise interpretation in physical systems \cite{vaccario2015first}.

\begin{figure}
    \centering
    \includegraphics[width=0.99\columnwidth, trim = 0cm 2.5cm 0.5cm 1.5cm, clip]{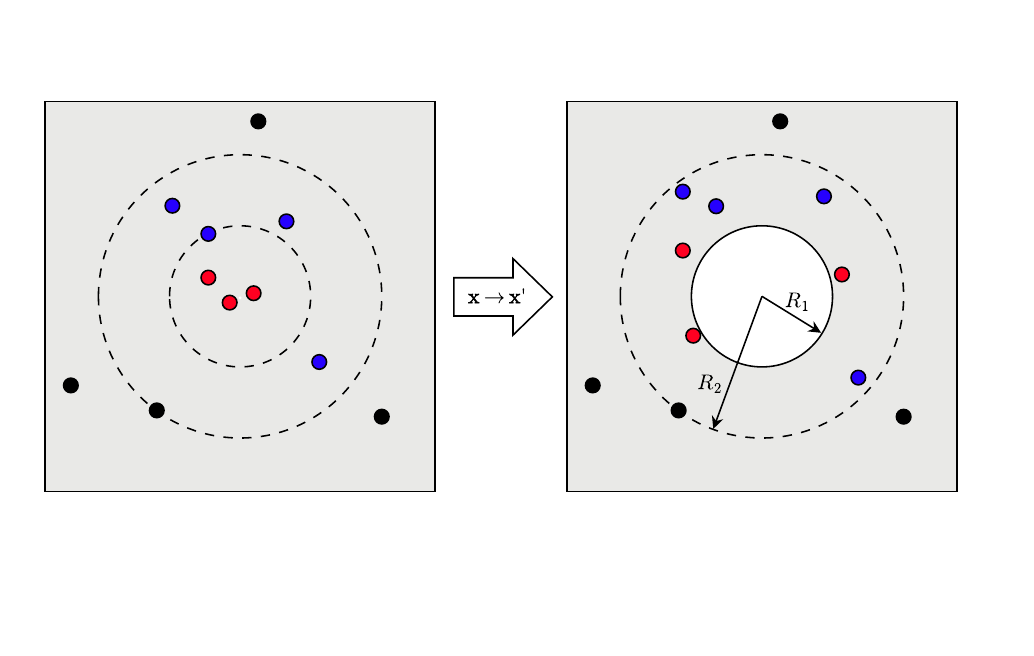}
    \caption{Schematic representation of a coordinate transformation that tears a hole in the metric space of a two-dimensional plane (left) to produce a finite region, indicated in white, that particles are unable to enter (right). Particle positions at radii $r \geq R_2$ (black points) are invariant under the transformation, while the radii of particles at $r < R_2$ (red and blue points) increase under the transformation.}
    \label{fig:CloakingSchematic}
\end{figure}

Despite these difficulties, we demonstrate that an annular metamaterial of spatially varying diffusivity can cloak an inner core from stochastically diffusing particles when interpreted under the It\^{o} convention. The properties of the metamaterial, or ``cloak'', are determined from a coordinate transformation that maps the inner core to the surrounding cloak, as in Fig.~\ref{fig:CloakingSchematic}.

In analogy to optical and thermal cloaking, ``stochastic'' cloaking prevents particles from accessing the core, while preserving the particle density outside the cloak. The latter is quantified by analysing the arrival distribution of particles downstream of the cloak.

\paragraph*{Langevin equation.}
The time evolution of a particle's position $x(t)$ in one dimension is described by the stochastic Langevin equation,
\begin{equation}\label{eq:LangevinEquation}
    \frac{m}{\gamma(x)}\ddot{x}(t) = - \dot{x}(t) + \sqrt{2 D(x)} \xi(t),
\end{equation}
where the dots above $x$ denote time derivatives, $m$ is the particle's mass, $\gamma(x)$ is the friction coefficient, $D(x)$ is the diffusivity, and $\xi(t)$ is Gaussian white noise with ensemble-averaged mean $\langle \xi(t) \rangle = 0$ and correlation $\langle \xi(t) \xi(t') \rangle = \delta(t-t')$. 
In the overdamped regime, where the typical timescale $\tau_o$ of interest satisfies $\tau_o \gg m/\gamma$, the Langevin equation simplifies to
\begin{equation}\label{eq:LangevinEquationOverdamped}
    \dot{x}(t) = \sqrt{2 D(x)} \xi(t),
\end{equation}
indicating the particle's dynamics are determined solely by $D(x)$.

\paragraph*{Conventions.}
To simulate a particle trajectory, one can iteratively integrate Eq.~(\ref{eq:LangevinEquationOverdamped}) in small timesteps $\Delta t$,
\begin{equation}\label{eq:LangevinEquationDiscrete}
    x(t + \Delta t) = x(t) + \int_t^{t+\Delta t} ds~\sqrt{2 D( x(s)) } \xi(s),
\end{equation}
which presents a choice when it comes to evaluating the integral on the right-hand side of Eq.~(\ref{eq:LangevinEquationDiscrete}). Specifically, we must choose which value the random variable $x(t)$ takes in the interval $[x(t), x(t+\Delta t)]$ in order to evaluate the diffusivity. However, there is no uniquely determined point at which it \textit{should} be evaluated. Instead, the choice is a matter of the adopted convention, which is parameterised by the continuous variable $\alpha \in [0,1]$ in the following \cite{lau2007state},
\begin{equation}\label{eq:LangevinEquationConvention}
    \begin{split}
        &x(t + \Delta t) = x(t)\\
        &+ \sqrt{2 D\left[ (1-\alpha)x(t) + \alpha x(t + \Delta t) \right]} \int_t^{t+\Delta t} ds~\xi(s),
    \end{split}
\end{equation}
where the integral $\int_t^{t+\Delta t} ds~\xi(s)$ is numerically evaluated by drawing from a Gaussian distribution of variance $\Delta t$.

Three common conventions are It\^{o} ($\alpha = 0$) \cite{ito1944109}, Stratonovich ($\alpha = 1/2$) \cite{stratonovich1966new}, and isothermal/H\"{a}nggi-Klimontovich ($\alpha=1$) \cite{hanggi1978stochastic,klimontovich1994nonlinear}, the latter of which we will refer to as ``anti-It\^{o}". Each convention has its merits \cite{lau2007state, escudero2023versus, sokolov2010ito}. However, integration under any $\alpha > 0$ convention invalidates the use of the standard Euler-Maruyama method of It\^{o} calculus because $\alpha > 0$ requires evaluating the diffusivity at a future timestep. The simplest way to simulate $\alpha > 0$ is thus to reformulate the dynamics in terms of the non-anticipatory It\^{o} convention, $\alpha = 0$, via a correction term (if sufficiently regular) \cite{lau2007state} or an auxiliary step \cite{perez2010stochastic, sagues2007spatiotemporal}---see Sec.~\ref{SM:SimulationScheme}. Importantly, the chosen convention affects the statistics of the particle's dynamics, as elucidated by the Fokker-Planck equation describing the time evolution of the particle density, 
\begin{equation}\label{eq:FokkerPlanckEquation}
    \frac{\partial P(x,t)}{\partial t} = \frac{\partial}{\partial x} D^{\alpha}(x) \frac{\partial}{\partial x} D^{1-\alpha}(x) P(x,t).
\end{equation}
Though similar to a heat equation, the Fokker-Planck equation differs by having an associated Langevin equation that directly governs the motion of each individual particle. However, due to their similarity, we will leverage methods from transformation thermodynamics \cite{guenneau2012transformation, schittny2013experiments} to inform our derivation of the spatially varying diffusivity that achieves stochastic particle cloaking in two dimensions. This diffusivity will then be used to simulate an ensemble of diffusive particles via the two-dimensional Langevin equation,
\begin{equation}\label{eq:LangevinEquation2D}
    \begin{split}
        &x_i(t + \Delta t) = x_i(t)\\
        &+ g_{ij}[ (1-\alpha)\mathbf{x}(t) + \alpha \mathbf{x}(t + \Delta t) ] \int_t^{t+\Delta t} ds~\xi_j(s),
    \end{split}
\end{equation}
where $i,j \in \{1,2\}$ index the components of vectors (such as the position $\mathbf{x} = (x_1,x_2)$ and noise $\bm{\xi} = (\xi_1,\xi_2)$), and matrices (such as the square root of the diffusivity/diffusion tensor $\mathsf{g} = \sqrt{2\mathsf{D}} = ((g_{11}, g_{21})^{\mathrm{T}}, (g_{12}, g_{22})^{\mathrm{T}})$). Above and throughout, repeated indices imply summation. We discuss the feasibility of generalising our results to a three-dimensional spherical cloak in Sec.~\ref{SM:3D-DiffusionTensor}.

\paragraph*{Transformed Fokker-Planck equation.}
To derive the cloak's spatially varying diffusivity, we consider a coordinate transformation that maps the protected circular core of radius $R_1$ to a surrounding annulus $R_1 \leq r' \leq R_2$, see Fig.~\ref{fig:CloakingSchematic}, where $r'$ is the radial distance from the centre of the core in the new coordinates. This annular region acts as the cloak, with its diffusivity determined through the specific mapping.

First, we consider the effect of a general coordinate transformation $\mathbf{x} \rightarrow \mathbf{x}'$ on the Fokker-Planck equation in two dimensions. Starting from the Fokker-Planck equation describing homogeneous diffusion $D_0$,
\begin{equation}\label{eq:FokkerPlanck2DUniform}
    \frac{\partial P(\mathbf{x},t)}{\partial t} = \frac{\partial}{\partial x_i} D_0 \frac{\partial P(\mathbf{x},t)}{\partial x_i},
\end{equation}
we seek the transformed diffusion tensor $\mathsf{D}'$ in the new coordinates that renders the Fokker-Planck equation form-invariant. In other words, by manipulating the transformed Fokker-Planck equation into the same form as Eq.~(\ref{eq:FokkerPlanck2DUniform}), identification of $\mathsf{D}'$ in place of $D_0$ readily follows. Leaving the details to Sec.~\ref{SM:TransformedFokkerPlanck}, we have for the Fokker-Planck equation after a general coordinate transformation \cite{guenneau2012transformation},
\begin{equation}\label{eq:FokkerPlanckTransformed}
    \frac{\partial P(\mathbf{x}',t)}{\partial t} = \frac{1}{\det(\mathsf{J})}\frac{\partial}{\partial x'_j} J^{-1}_{kj} D_0 J^{-1}_{kl}\det(\mathsf{J})\frac{\partial P(\mathbf{x}',t)}{\partial x'_l},
\end{equation}
where the Jacobian matrix $\mathsf{J}$ has components $J_{ij} = \partial_{x'_j}x_i$. Equation~(\ref{eq:FokkerPlanckTransformed}) poses the dynamics on an effective manifold \cite{hsu2002stochastic}, whereas we desire the coordinate transformation to leave the Fokker-Planck equation in an invariant ``Euclidean'' form analogous to Eq.~(\ref{eq:FokkerPlanck2DUniform}). However, this is prevented by the space-dependent factor $1/\det (\mathsf{J})$ preceding the spatial derivatives on the right-hand side of Eq.~(\ref{eq:FokkerPlanckTransformed}). For the heat equation in transformation thermodynamics, this factor is merely absorbed into the specific heat capacity \cite{schittny2013experiments, guenneau2012transformation}. However, there is no analogous trick for the case of stochastic particle diffusion here, since it depends solely on the diffusivity $\mathsf{D}'(\mathbf{x})$. Hence, unlike the heat equation, the Fokker-Planck equation is generally not form-invariant under a coordinate transformation, as Eq.~(\ref{eq:FokkerPlanckTransformed}) is the furthest one can simplify before considering specific transformations.

\paragraph*{Non-linear transformation.}
Equation~(\ref{eq:FokkerPlanckTransformed}) suggests a transformation with constant $\det(\mathsf{J})$ would render the Fokker-Planck equation invariant. The most-studied transformation is the linear mapping $r' = R_1 + r (R_2 - R_1) / R_2$ for $r \leq R_2$ \cite{pendry2006controlling, greenleaf2003anisotropic, greenleaf2003nonuniqueness}. This has non-constant $\det(\mathsf{J}) = (R_2/(R_2-R_1))^2 (r' - R_1)/r'$ for $r \leq R_2$, which motivates us to consider alternative transformations. As it turns out, a transformation that simultaneously maps the protected region, $r' < R_1$, to the annulus, $R_1 \leq r' \leq R_2$, while maintaining constant $\det(\mathsf{J})$ is difficult to find. \textit{Close} to satisfying both properties is the non-linear transformation \cite{qiu2009spherical, guenneau2017cloaking},
\begin{equation}\label{eq:NonLinearTransformation}
    r' =
    \begin{cases}
        \sqrt{\beta r^2 + R_1^2},& \quad r \leq R_2,\\
        r,& \quad r > R_2,
    \end{cases}
\end{equation}
where $\beta = (R_2^2 - R_1^2)/R_2^2$ and the azimuthal coordinate $\varphi' = \varphi$ is invariant. In Cartesian coordinates $\mathbf{x}' = (x',y')$, the Jacobian for this non-linear transformation is
\begin{equation}\label{eq:NonLinearTransformation_Jacobian}
\begin{split}
    \mathsf{J}
    &= \frac{\partial (x,y)}{\partial (r,\varphi)}\frac{\partial (r,\varphi)}{\partial (r',\varphi')}\frac{\partial (r',\varphi')}{\partial (x',y')}\\
    &= 
    \begin{cases}
        \mathsf{R}(\varphi') \mathrm{diag}\left( \frac{r'}{r \beta}, \frac{r}{r'}\right) \mathsf{R}(-\varphi'), & \quad R_1 < r' \leq R_2,\\
        \mathbbm{1}, & \quad r' > R_2,
    \end{cases}
\end{split}
\end{equation}
where $\mathsf{R}(\varphi)$ and $\mathbbm{1}$ are the $2\times 2$ rotation and identity matrices, respectively. The Jacobian (\ref{eq:NonLinearTransformation_Jacobian}) has a piecewise-constant determinant,
\begin{equation}\label{eq:NonLinearTransformation_detJ}
    \det(\mathsf{J})
    = 
    \begin{cases}
        \frac{1}{\beta}, & \quad R_1 < r' \leq R_2,\\
        1, & \quad r' > R_2,
    \end{cases}
\end{equation}
suggesting Eq.~(\ref{eq:FokkerPlanckTransformed}) can be brought into the desired form (\ref{eq:FokkerPlanck2DUniform}) on a piecewise basis under the transformation (\ref{eq:NonLinearTransformation}) above. On this piecewise basis, the factors of $\det(\mathsf{J})$ in Eq.~(\ref{eq:FokkerPlanckTransformed}) cancel, allowing us to approximate the Fokker-Planck equation under this transformation by
\begin{equation}\label{eq:FokkerPlanckTransformed_Final}
    \frac{\partial P(\mathbf{x}',t)}{\partial t} \approx \frac{\partial}{\partial x'_j} \mathsf{D}'_{jk}\frac{\partial P(\mathbf{x}',t)}{\partial x'_k},
\end{equation}
where we have identified the piecewise diffusion tensor,
\begin{widetext}
\begin{equation}\label{eq:TransformedDiffusionTensor}
    \mathsf{D}'
    = 
    D_0 \mathsf{J}^{-\mathrm{T}}\mathsf{J}^{-1}
    =
    \begin{cases}
        D_0 \beta \mathsf{R}(\varphi') \mathrm{diag}\left(\frac{r'^2 - R_1^2}{r'^2}, \frac{r'^2}{r'^2- R_1^2} \right) \mathsf{R}(-\varphi'), & \quad R_1 < r' \leq R_2,\\
        D_0 \mathbbm{1}, & \quad r' > R_2,
    \end{cases}
\end{equation}
\end{widetext}
from which we can recover homogeneous diffusion $\mathsf{D}' = D_0 \mathbbm{1}$, as in Eq.~(\ref{eq:FokkerPlanck2DUniform}), by setting $R_1 = 0$. Strictly speaking, Eq.~(\ref{eq:FokkerPlanckTransformed_Final}) is an equality everywhere except at $r' = R_2$, where there is an additional divergent contribution arising from the sharp jump in $\det(\mathsf{J})$, Eq.~(\ref{eq:NonLinearTransformation_detJ}). By neglecting this term, we na\"{i}vely expect the diffusion tensor in Eq.~(\ref{eq:TransformedDiffusionTensor}) to yield far-from-perfect cloaking. Furthermore, Eq.~(\ref{eq:FokkerPlanckTransformed_Final}) has the appearance of a two-dimensional analogue to the one-dimensional Fokker-Planck equation (\ref{eq:FokkerPlanckEquation}) with $\alpha = 1$, na\"{i}vely suggesting the anti-It\^{o} convention \cite{hanggi1978stochastic, klimontovich1994nonlinear} of the Langevin equation (\ref{eq:LangevinEquation2D}) would yield the best cloaking performance. However, such a prediction is based on the misguided presumption that the identification of $\alpha$ from Eq.~(\ref{eq:FokkerPlanckTransformed_Final}) equally applies to Fokker-Planck equations other than just the one-dimensional case \cite{lau2007state}. In fact, in higher dimensions, the correction term that arises when treating anti-It\^{o} as a perturbation to It\^{o} is generally not of the divergence form necessary to write the Fokker-Planck equation as a two-dimensional analogue of Eq.~(\ref{eq:FokkerPlanckEquation}) \cite{escudero2023versus}. As a result, we also trialled the diffusion tensor $\mathsf{D}'$, Eq.~(\ref{eq:TransformedDiffusionTensor}), in the Langevin equation (\ref{eq:LangevinEquation2D}) for conventions other than just anti-It\^{o}. This was further justified \textit{a posteriori}, since we found the It\^{o} convention \cite{ito1944109} resulted in the best cloaking performance. This is likely due to It\^o requiring less regularity than the other conventions \cite{correales2019ito,escudero2020kinetic,escudero2023fluctuation,escudero2023versus}, which have singular correction terms in this case. This singular behaviour means the usual relation between the Langevin and Fokker-Planck equations is not guaranteed. Thus, the manipulations above can be considered only formal steps at first, becoming effectively valid (or not) only after appealing to simulations.

\begin{figure*}
    \centering
    \includegraphics[width=0.99\textwidth, trim = 4cm 0cm 1.5cm 0cm, clip]{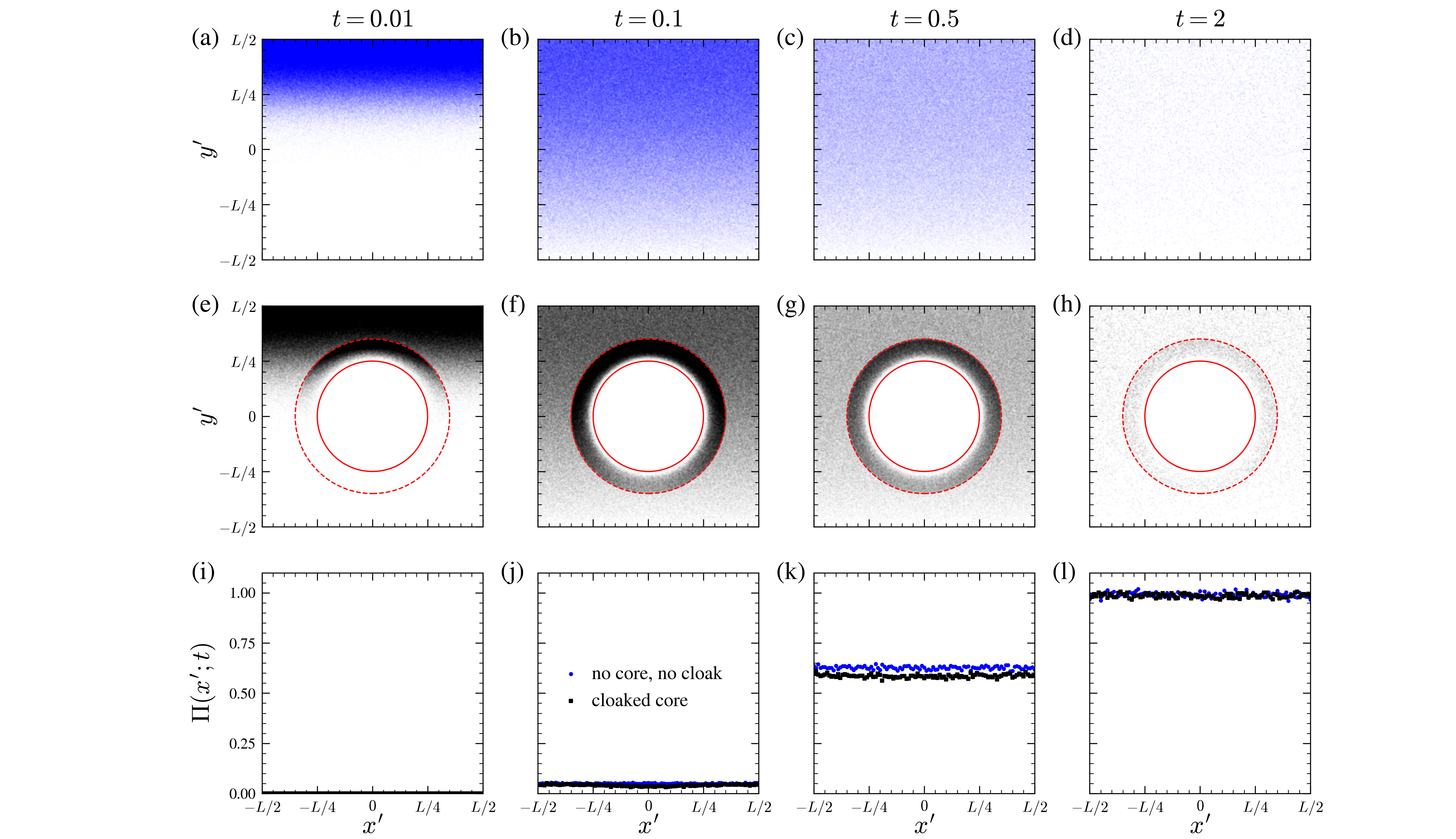}
    \caption{Comparison between simulations of no core (a)--(d), and a cloaked core (e)--(h), with their corresponding arrival distributions $\Pi(x';t)$, Eq.~(\ref{eq:ArrivalDistribution}), (i)--(l) at times $t = 0.01$ in (a), (e), and (i); $t = 0.1$ in (b), (f), and (j); $t=0.5$ in (c), (g), and (k); and $t=2$ in (d), (h), and (l). Each point in (a)--(h) represents the position of one of $N = 10^6$ particles. The simulation parameters used in all subfigures were $\Delta t = 10^{-5}$, $L=1$, $D_0 = 1$. In (e)--(h), the inner core has radius $R_1 = 0.25$ (solid red line) and is surrounded by an annular cloak of outer radius $R_2 = 0.35$ (dashed red line). Particles are prevented from penetrating the core by the cloak in (e)--(h), and the arrival distributions $\Pi(x';t)$ in (i)--(l) show that of the cloaked core closely matches that of no core for all times $t$, suggesting the diffusion tensor, Eq.~(\ref{eq:TransformedDiffusionTensor}), used in the cloaking simulations produces near-perfect cloaking.}
    \label{fig:TimeGrid}
\end{figure*}

\paragraph*{Simulation setup.} We performed simulations to determine how well the cloak, with diffusivity given by Eq.~(\ref{eq:TransformedDiffusionTensor}), conceals the inner core. The core of radius $R_1$ is centred in a square box of linear size $L$. The box has periodic boundary conditions at the sides, $x' = -L/2$ and $x' = L/2$, a reflecting boundary at the top, $y' = L/2$, and an absorbing boundary at the bottom, $y' = -L/2$, through which all particles eventually leave the system. A uniform line of $N$ particles is initialised at the top, $y' = L/2$, at time $t=0$, i.e.\ $P(x',y';t=0) = \delta(y'-L/2)\sum_{i=1}^{N} \delta(x' + L/2 - iL/N)/N$. This setup is notably different from that of thermal cloaking, where the temperatures at the top and bottom of the box are kept fixed. Here, the setup is analogous to a single pulse of radiation emitted from the line $y' = L/2$ at $t=0$. Strictly speaking, each particle contributes a Dirac delta function to the overall particle density $P(x',y';t)$. However, for finite $N$, we rather view the particle density in a coarse-grained sense, i.e.\ envisage the simulation box as being divided into small cells of linear length $\ell$, such that the density in a particular cell is well approximated by $n/(N \ell^2)$, where $n$ is the number of particles contained in that cell. In this sense, the true particle density $P(x',y';t)$ is recovered in the thermodynamic limit $N \to \infty$, while also taking $\ell \to 0$.

After initialisation, particles diffuse according to Eq.~(\ref{eq:LangevinEquation2D}) with diffusivity given by Eq.~(\ref{eq:TransformedDiffusionTensor})---see Sec.~\ref{SM:SimulationScheme} for full details. In case particles penetrated the core $r' \leq R_1$, we chose it to have the same homogeneous diffusivity $D_0$ as the medium surrounding the cloak. Perfect cloaking would prevent particles from penetrating the core, $r' \leq R_1$, while also ensuring the density outside the cloak, $r' > R_2$, matches that of a simulation containing no core at all times $t$. In other words, any observer measuring the density at a radius $r' > R_2$ would be unable to detect the presence of the core \textit{and} the cloak because the density would be the same as in their absence. How closely these densities match gives a measure of the cloak's performance. To quantify cloaking performance, we introduced the cumulative arrival distribution $\Pi(x,t_f)$ of particles at the absorbing boundary, which is analogous to a ``splitting probability'' in a first-passage problem \cite{van1992stochastic},
\begin{equation}\label{eq:ArrivalDistribution}
    \Pi(x';t_f) = \int_0^{t_f} dt~D_0 \left. \frac{\partial P(x',y';t)}{\partial y'} \right|_{y'=-L/2},
\end{equation}
where $D_0 \partial_y P(x', y'; t)$ is the particle current at $\mathbf{x}' = (x',y')$ in the negative-$y'$ direction. Equation~(\ref{eq:ArrivalDistribution}) is the density of particles that have arrived at position $x'$ on the absorbing boundary by time $t_f$. The spatial average of $\Pi(x';t_f)$ asymptotically approaches unity as $t_f \to \infty$, i.e.\ $\lim_{t_f \to \infty}\int_{-L/2}^{L/2} dx'~ \Pi(x';t_f)/L = 1$, signifying the absorption of all particles. For the cloaking to be considered effective, the arrival distribution in the simulation of the cloaked core must closely match that of a simulation of no core (and no cloak).

\paragraph*{Simulation results.}
We performed the simulations described above for the It\^{o} ($\alpha = 0$) \cite{ito1944109}, Stratonovich ($\alpha = 1/2$) \cite{stratonovich1966new}, and anti-It\^{o} ($\alpha = 1$) \cite{hanggi1978stochastic, klimontovich1994nonlinear} conventions of interpreting the Langevin equation~(\ref{eq:LangevinEquation2D}). The Stratonovich and anti-It\^{o} conventions resulted in poorer cloaking than that of It\^{o}, signified by: particles penetrating the core; spurious large jumps near $r'=R_1$; and noticeable differences in the arrival distributions, Eq.~(\ref{eq:ArrivalDistribution}), compared to that of no core. This is not surprising given the application of these conventions to systems involving ``problematic'' boundaries---like the discontinuity at $r'=R_2$ and singularity at $r'=R_1$ in the diffusivity, Eq.~(\ref{eq:TransformedDiffusionTensor})---can lead to ill-posed results, as mentioned in Refs.~\cite{escudero2023versus, correales2019ito, escudero2020kinetic, escudero2023fluctuation}. Given their relatively poor performance, we omit further discussion of the Stratonovich and anti-It\^{o} conventions.

Our key result is that, remarkably, the It\^{o} convention exhibited near-perfect cloaking signified by no particle penetration of the core and an arrival distribution, Eq.~(\ref{eq:ArrivalDistribution}), that closely matched that of no core for all times $t$, see Fig.~\ref{fig:TimeGrid}. ``Near perfect'' is to caveat a small discrepancy in the number of particles that have been absorbed up to time $t$ between the simulations of the cloaked core and no core. However, this is similar to time lags typically seen in thermal cloaking \cite{craster2018cloaking}. Moreover, this discrepancy all but disappears through regularisation procedures that we introduce in Sec.~\ref{SM:regularisation}. Strikingly, the particle density inside the cloak $R_1 < r' \leq R_2$ is markedly different from that seen for thermal cloaking, where the density typically decays smoothly from the outer region $r' > R_2$ to the inner cloak boundary $r' = R_1$ \cite{guenneau2012transformation}. Here, there is a distinct ``halo'' of high particle density in the cloak to accommodate those particles that would otherwise be inside the core $r' < R_1$. This also suggests why the smooth decay in density for thermal cloaking comes at the cost of penetration of the core \cite{guenneau2012transformation}.

\paragraph*{Discussion and conclusion.}
Inspired by optical and thermal cloaking \cite{pendry2006controlling, guenneau2012transformation}, we demonstrated ``stochastic cloaking'' of a region from diffusive particles. This is in spite of the lack of form-invariance of the Fokker-Planck equation under a general coordinate transformation, as well as the inherent difficulties introduced by stochasticity. Our setup involved a circular core that we attempted to conceal by surrounding it with an annular metamaterial of spatially varying diffusivity. To quantify the cloaking performance, we measured the arrival distribution, Eq.~(\ref{eq:ArrivalDistribution}), of particles along a tangential line downstream of the cloak. The hallmark of cloaking in this case is an arrival distribution that matches that of a simulation involving no core for all times. Our key result is a demonstration of near-perfect cloaking for the diffusion tensor $\mathsf{D}'$ in Eq.~(\ref{eq:TransformedDiffusionTensor}) if the Langevin equation is interpreted under the It\^{o} convention \cite{ito1944109}, see Fig.~\ref{fig:TimeGrid}. The cloaking performance can be further improved through regularisation procedures introduced in Sec.~\ref{SM:regularisation}. These procedures also soften the requirements for the cloak to be implemented in an experiment by removing any singular behaviour. While we demonstrated near-perfect cloaking, the specific coordinate transformation we considered resulted in invariance of the Fokker-Planck equation only on a piecewise basis. Therefore, future work should focus on establishing a firmer theoretical basis for the cloaking that results from the It\^{o} convention. Another interesting avenue would be to test stochastic cloaking on other cloak geometries, such as diamonds \cite{li2009near}.

\begin{acknowledgments}
C.R.\ acknowledges support from the Engineering and Physical Sciences Research Council (Grant No.\ 2478322). C.E.\ and H.R.\ acknowledge support from the Government of Spain (Ministerio de Ciencia e Innovación) through Project No.~PID2021-125871NB-I00.
\end{acknowledgments}

\bibliography{references}




\balancecolsandclearpage
\onecolumngrid

\section*{Supplementary Material}

\renewcommand{\theequation}{S\arabic{equation}}
\setcounter{equation}{0}

\renewcommand{\thefigure}{S\arabic{figure}}
\setcounter{figure}{0}

\renewcommand{\thesection}{S\Roman{section}}
\setcounter{section}{0}


\section{Coordinate transformation of the Fokker-Planck equation} \label{SM:TransformedFokkerPlanck}

In this supplementary section, we derive Eq.~(8) of the main text, i.e.\ the result of applying a coordinate transformation to the Fokker-Planck equation (7) describing homogeneous diffusion of strength $D_0$. For convenience, we repeat the latter here,
\begin{equation}\label{eq:FokkerPlanck2DUniform_Appendix}
    \frac{\partial P(\mathbf{x},t)}{\partial t} = \frac{\partial}{\partial x_i} D_0 \frac{\partial P(\mathbf{x},t)}{\partial x_i},
\end{equation}
where, as in the main text, a repeated index implies summation over that index.

The standard way \cite{guenneau2012transformation} to proceed is to integrate Eq.~(\ref{eq:FokkerPlanck2DUniform_Appendix}) against a scalar ``test function'' $\Phi(\mathbf{x})$, where $\mathbf{x} = (x_1,x_2)$, whose properties are arbitrary aside from being infinitely differentiable and having compact support on the region $\Omega(\mathbf{x}) = [-L/2,L/2] \times [-L/2,L/2]$ in which the particles reside. From integrating Eq.~(\ref{eq:FokkerPlanck2DUniform_Appendix}), we obtain
\begin{equation}\label{eq:FokkerPlanckTestFunction_IntParts}
    \int_{\Omega(\mathbf{x})} dx_1 dx_2~\frac{\partial P(\mathbf{x},t)}{\partial t}\Phi(\mathbf{x}) = -\int_{\Omega(\mathbf{x})} dx_1 dx_2~\frac{\partial \Phi(\mathbf{x})}{\partial x_i} D_0 \frac{\partial P(\mathbf{x},t)}{\partial x_i},
\end{equation}
where the boundary term from the integration by parts on the right-hand side vanishes due to the compact support property of $\Phi(\mathbf{x})$.

Now, we consider the effect of a general coordinate transformation $\mathbf{x} = (x_1,x_2) \to \mathbf{x}' = (x'_1, x'_2)$. Under the change of coordinates, all instances of $\partial_{x_j}$ in Eq.~(\ref{eq:FokkerPlanckTestFunction_IntParts}) are replaced with $J_{ij}^{-1} \partial_{x'_j}$, where $J_{ij} = \partial_{x'_j}x_i$ are elements of the Jacobian matrix $\mathsf{J}$, and $dx_1 dx_2 = \det(\mathsf{J}) dx'_1 dx'_2$, i.e.\
\begin{equation}\label{eq:FokkerPlanckTestFunction_Transform}
    \int_{\Omega(\mathbf{x}')} dx'_1 dx'_2~\det(\mathsf{J}) \frac{\partial P(\mathbf{x}',t)}{\partial t}\Phi(\mathbf{x}') = -\int_{\Omega(\mathbf{x}')} dx'_1 dx'_2~J_{ji}^{-1}\frac{\partial \Phi(\mathbf{x}')}{\partial x'_i} D_0 J_{jk}^{-1} \det(\mathsf{J}) \frac{\partial P(\mathbf{x}',t)}{\partial x'_k}.
\end{equation}

The integrand on the right-hand side of Eq.~(\ref{eq:FokkerPlanckTestFunction_Transform}) is merely the scalar product of the two vectors $\partial_{x'_i}\Phi(\mathbf{x}')$ and $J_{ji}^{-1} D_0 J_{jk}^{-1} \det(\mathsf{J}) \partial_{x'_k}P(\mathbf{x}',t)$. Hence, upon another integration by parts to return the derivative $\partial_{x'_j}$ acting on the former to acting on the latter, followed by some rearrangement, we find
\begin{equation}\label{eq:FokkerPlanckTestFunction_Final}
    \int_{\Omega(\mathbf{x}')} dx'_1 dx'_2~\Phi(\mathbf{x}') \left(\det(\mathsf{J}) \frac{\partial P(\mathbf{x}',t)}{\partial t} - \frac{\partial }{\partial x'_i} J_{ji}^{-1} D_0 J_{jk}^{-1} \det(\mathsf{J}) \frac{\partial P(\mathbf{x}',t)}{\partial x'_k} \right) = 0,
\end{equation}
where the boundary term resulting from the integration by parts once again vanishes. It follows that if $P(\mathbf{x}',t)$ satisfies Eq.~(\ref{eq:FokkerPlanckTestFunction_Final}) for an arbitrary test function $\Phi(\mathbf{x}')$, then it also satisfies Eq.~(8).

\section{Simulations of different discretisation conventions of the Langevin equation}\label{SM:SimulationScheme}

Here, we detail how the simulations of the different discretisation conventions of the Langevin equation were performed. As discussed in the main text, the most direct route to simulating the $\alpha > 0$ cases is by reformulating the numerical integration of the Langevin equation (6) in terms of the non-anticipatory It\^{o} convention \cite{ito1944109}, $\alpha = 0$, either through a correction term \cite{lau2007state} or an auxiliary step \cite{perez2010stochastic, sagues2007spatiotemporal}. The correction term that arises when treating the other conventions as a perturbation to It\^{o} typically involves spatial derivatives of the diffusion tensor $\mathsf{D}'$. Since the diffusion tensor here, Eq.~(13), has a jump discontinuity at $r' = R_2$, this makes it difficult to simulate the dynamics through the correction-term approach. Thus, we instead opted for the auxiliary-step approach, which is able to handle discontinuities \cite{perez2010stochastic}. Qualitatively, this approach involves first generating an auxiliary position $\mathbf{\tilde{x}}(t + \Delta t)$ at the future timestep under the It\^{o} convention. This auxiliary position is then interpreted as the ``future'' timestep at which to evaluate the diffusion tensor for an $\alpha > 0$ simulation, but still treating the Langevin equation (6) under the It\^{o} convention. This correctly recovers the statistics of an $\alpha > 0$ simulation \cite{perez2010stochastic, sagues2007spatiotemporal}. More explicitly, the simulations are implemented by the following pseudocode, where all positions are to be understood as those in the new coordinates \textit{after} applying the coordinate transformation:
\begin{small}
\SetKwInput{Kw}{initialise}
\begin{algorithm}[H]
 \Kw{ \\
 \quad  $ \bullet \ $ Set particle index: $n=1$;}

\While{$n \leq N$}{
 \Kw{ \\
 \quad  $ \bullet \ $ Set initial conditions: $t=0$, $y(0) = L/2$, $x(0)=n L / N - L/2$;}

 \While{$t<t_f \mathrm{~and~} y(t) > -L/2$}{
 
  $\bullet \ $Generate two independent random numbers, $\mathcal{N}_1(0,\Delta t)$ and $\mathcal{N}_2(0,\Delta t)$, drawn from a Gaussian distribution of mean $0$ and variance $\Delta t$, i.e.\ a distribution $G(x) = \exp(-x^2/(2\Delta t))/\sqrt{2 \pi \Delta t}$\;
  
  $\bullet \ $ Calculate auxiliary ``It\^{o}'' position:\\
  $\tilde{x}_I(t + \Delta t) = x(t) + g_{11}[\mathbf{x}(t)]\mathcal{N}_1(0,\Delta t) + g_{12}[\mathbf{x}(t)] \mathcal{N}_2(0,\Delta t) $\;
  $\tilde{y}_I(t + \Delta t) = y(t) + g_{21}[\mathbf{x}(t)]\mathcal{N}_1(0,\Delta t) + g_{22}[\mathbf{x}(t)]\mathcal{N}_2(0,\Delta t) $\;

  $\bullet \ $ Calculate $\alpha$-dependent position:\\

  \If{$0 < \alpha \leq 1$}{
  $x(t + \Delta t) = x(t) + g_{11}[(1 - \alpha)\mathbf{x}(t) + \alpha\mathbf{\tilde{x}}_I(t + \Delta t)]\mathcal{N}_1(0,\Delta t) + g_{12}[(1 - \alpha)\mathbf{x}(t) + \alpha\mathbf{\tilde{x}}_I(t + \Delta t)] \mathcal{N}_2(0,\Delta t) $; \\
  $y(t + \Delta t) = y(t) + g_{21}[(1 - \alpha)\mathbf{x}(t) + \alpha\mathbf{\tilde{x}}_I(t + \Delta t)]\mathcal{N}_1(0,\Delta t) + g_{22}[(1 - \alpha)\mathbf{x}(t) + \alpha\mathbf{\tilde{x}}_I(t + \Delta t)] \mathcal{N}_2(0,\Delta t) $;
  }

  \Else{
  \If{$\alpha = 0$}{
  $x(t + \Delta t) = \tilde{x}_I(t + \Delta t) $; \\
  $y(t + \Delta t) = \tilde{y}_I(t + \Delta t) $;
  }
  }

  $\bullet \ $ Implement periodic boundary conditions:\\
  
  \If{$x(t+\Delta t) \geq L/2$ }{
   $\bullet \ $ $x(t+\Delta t) = x(t+\Delta t) - L$;
   }
   \If{$x(t+\Delta t) < L/2$ }{
   $\bullet \ $ $x(t+\Delta t) = x(t+\Delta t) + L$;
   }

   $\bullet \ $ Implement reflecting boundary condition:\\
   \If{$y(t+\Delta t) > L/2$ }{
   $\bullet \ $ $y(t+\Delta t) = L - y(t+\Delta t)$;
   }
   
   $\bullet \ $ Increment time: $t = t + \Delta t$\;
}

$\bullet \ $ Increment particle number: $n = n + 1$;
 
}
\end{algorithm}
\end{small}
\noindent where $t_f$ is the final time that the simulation is run to for each particle, $\Delta t$ is the simulation timestep, $N$ is the total number of particles, and $g_{ij}$ are the elements of the matrix $\mathsf{g} = \sqrt{2\mathsf{D}}$, as in the main text, where $\mathsf{D}$ can represent either the singular diffusion tensor of Eq.~(13), or the non-singular diffusion tensor of Eq.~(\ref{eq:TransformedDiffusionTensor_NonSingular}), see SM Sec.~\ref{SM:regularisation}, depending on whichever diffusion tensor is being simulated.

\section{Regularisation and realistic cloaks}\label{SM:regularisation}

\begin{figure}
    \centering
    \includegraphics[width=0.91\columnwidth, trim = 2cm 0cm 1.5cm 1cm, clip]{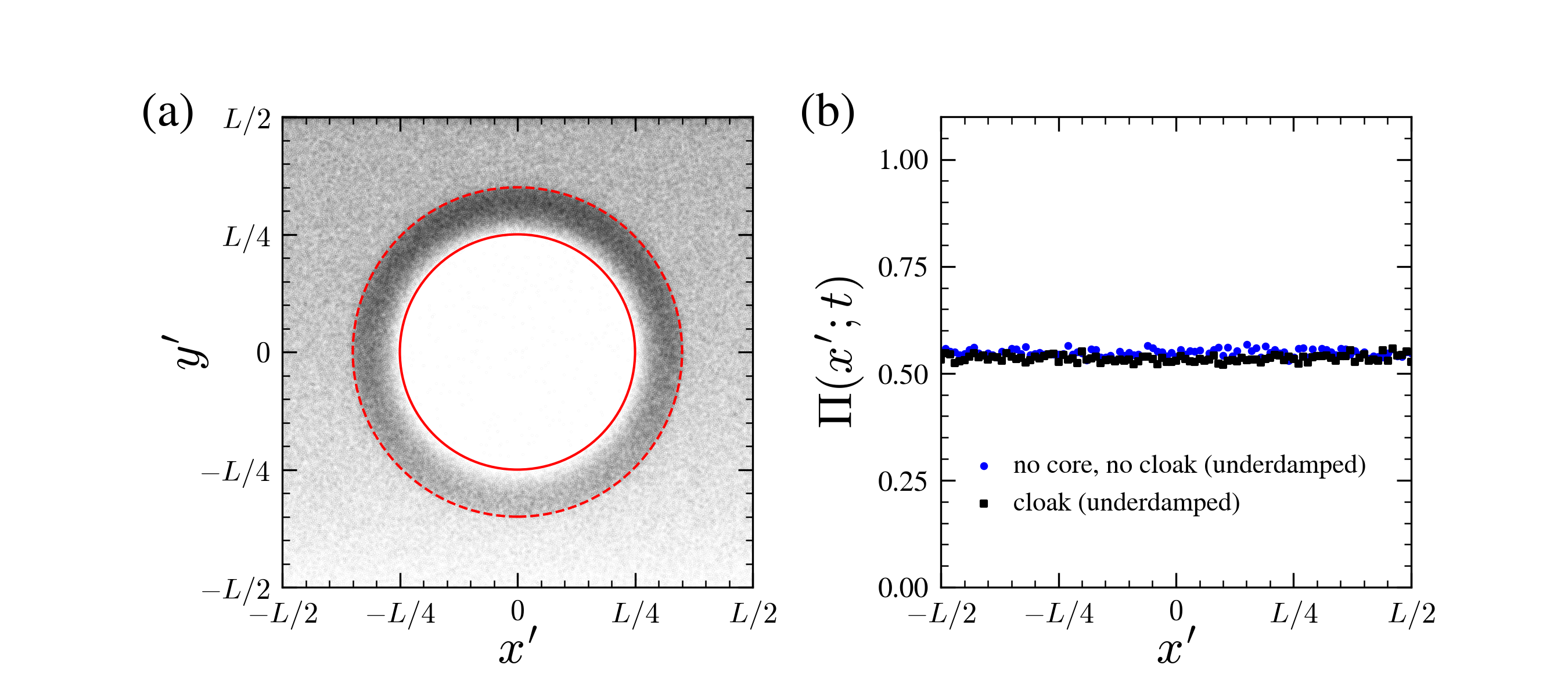}
    \caption{(a) Snapshot at time $t = 0.5$ from a simulation of the underdamped dynamics described by Eq.~(\ref{eq:LangevinEquationUnderdamped2D}), for constant $\gamma = 1$ and using the diffusion tensor given in Eq.~(13). Details of how the underdamped dynamics were simulated are given in Sec.~\ref{SM:UnderdampedSimulations}. Each data point in (a) represents the position of one of $N = 10^6$ particles. (b) Corresponding cumulative arrival distribution $\Pi(x;t)$, Eq.~(14), for the underdamped simulations of a cloaked core compared against that of no core. The simulation parameters used in both subfigures were $\Delta t = 10^{-5}$, $L=1$, $D_0 = 1$, $m = 10^{-3}$, $R_1 = 0.25$, and $R_2 = 0.35$. We considered a small mass $m$ such that the inertial timescale $m / \gamma \ll L^2 / D_0$, thus making the dynamics close to the overdamped regime, illustrated in Fig.~2. However, the fact that there is a finite mass $m$ is enough to regularise the singular behaviour of the diffusion tensor, Eq.~(13), as evidenced by the cumulative arrival distribution $\Pi(x;t)$ for the cloaked core now more closely matching that of no core compared to the overdamped case. Furthermore, compared to the overdamped simulations illustrated in Fig.~2, less particles have been absorbed at $y' = -L/2$ up to time $t = 0.5$ because some particles ``stick'' to the reflecting boundary $y' = L/2$ for short periods of time due to their inertia. We also observe there to be some (albeit a very small amount of) particle penetration of the inner core in the underdamped case, which is the price paid for the regularisation provided by the finite mass $m$.}
    \label{fig:UnderdampedSnapshot}
\end{figure}

To corroborate the findings in the main text of near-perfect cloaking under the It\^{o} convention, Fig.~2, we also performed simulations for two cases of stochastic cloaking that could more feasibly be implemented in an experiment, Secs.~\ref{SM:UnderdampedSimulations} and \ref{SM:NonSingularDiffusionTensor}. We also discuss in Sec.~\ref{SM:3D-DiffusionTensor} the possibility of extending the results of the main text to a three-dimensional spherical cloak.

\subsection{Underdamped dynamics}\label{SM:UnderdampedSimulations}

The first of these is for the more general underdamped dynamics described by the two-dimensional analogue of Eq.~(1), i.e.\
\begin{equation}\label{eq:LangevinEquationUnderdamped2D}
    m\ddot{\mathbf{x}}(t) = - \mathsf{\gamma}(\mathbf{x})\dot{\mathbf{x}} + \mathsf{\gamma}(\mathbf{x})\mathsf{g}(\mathbf{x}) \bm{\xi}(t),
\end{equation}
where the symbols retain their same definitions as in the main text. Here, the sharp jumps in the diffusivity $\mathsf{D}'$, Eq.~(13), are regularised by the particles having inertia. The results of the overdamped case in the main text can be recovered for the underdamped dynamics by taking particle mass $m \to 0$ for a spatially varying diffusivity but constant friction coefficient. This is consistent with the recipe to recover the It\^{o} convention of an overdamped Langevin equation from a discretisation-independent underdamped Langevin equation \cite{bo2017multiple, matsuo2000stochastic, yang2013brownian, stolovitzky1998non}. A snapshot of the particle density and arrival distribution from an underdamped simulation for constant $\gamma = 1$ (corresponding to $\alpha = 0$ in the overdamped regime) is illustrated in Fig.~\ref{fig:UnderdampedSnapshot}. As for the overdamped dynamics considered in the main text, Fig.~2, spatially varying $\gamma$ (corresponding to $\alpha = 1$ in the overdamped regime) in the underdamped dynamics yet again produced relatively poor cloaking. For the finite but small $m$ used in Fig.~\ref{fig:UnderdampedSnapshot}, the most notable difference to that of the overdamped dynamics, Fig.~2, is that the arrival distributions between the cloaked core and no core in the underdamped case agree more closely at all times than for the overdamped dynamics---though this comes at the price of some particle penetration of the inner core. The above results suggest that stochastic cloaking could be implemented in an experiment by tracking colloidal particles moving through a material of spatially varying temperature in a similar setup to that of Ref.~\cite{schittny2013experiments} for spatially varying conductivity. This could be further facilitated by the homogenisation procedure outlined in Ref.~\cite{guenneau2012transformation}, which demonstrates that the properties of a cloak with continuously varying parameters can be approximated by a multilayered cloak consisting of concentric shells of uniform and isotropic parameter values. In the present case, this would correspond to shells of constant temperature.

We now provide the pseudocode for the simulations of the underdamped Langevin equation (\ref{eq:LangevinEquationUnderdamped2D}). By Einstein's relation, diffusion $D$ satisfies $\gamma D = k_B T$, where $\gamma$ is the friction coefficient, $k_B$ is the Boltzmann constant, and $T$ is the temperature of the heat bath from which the diffusion derives. Hence, spatial modulation of the particle diffusivity can be achieved through either that of the friction $\gamma$ or the temperature $T$. In the limit of particle mass $m \to 0$, the former corresponds to an anti-It\^{o} convention, $\alpha = 1$, of the overdamped Langevin equation (6) \cite{sancho1982adiabatic, volpe2016effective, lau2007state}, while the latter corresponds to that of an It\^{o} convention, $\alpha = 0$ \cite{bo2017multiple, matsuo2000stochastic, yang2013brownian, stolovitzky1998non}. To allow a choice between the two, we implemented the underdamped dynamics through the following pseudocode:
\begin{small}
\SetKwInput{Kw}{initialise}
\begin{algorithm}[H]
 \Kw{ \\
 \quad  $ \bullet \ $ Set particle index: $n=1$\;
 \quad  $ \bullet \ $ Choose $\alpha \in \{0 , 1\}$;}

\While{$n \leq N$}{
 \Kw{ \\
 \quad  $ \bullet \ $ Set initial conditions: $t=0$, $y(0) = L/2$, $x(0)=n L / N - L/2$, $v_x(0) = 0$, $v_y(0) = 0$;}

 \While{$t<t_f \mathrm{~and~} y(t) > -L/2$}{
 
  $\bullet \ $Generate two independent random numbers, $\mathcal{N}_1(0,\Delta t)$ and $\mathcal{N}_2(0,\Delta t)$, drawn from a Gaussian distribution of mean $0$ and variance $\Delta t$, i.e.\ a distribution $G(x) = \exp(-x^2/(2\Delta t))/\sqrt{2 \pi \Delta t}$\;
  
  $\bullet \ $ Update velocity according to the Langevin equation (\ref{eq:LangevinEquationUnderdamped2D}):\\
  \If{$\alpha = 1$ }{
   $\bullet \ $ Friction term is spatially varying, i.e.\\
   $v_x(t+\Delta t) = v_x(t) + \frac{1}{m}\big[ -v_x(t) \Delta t (\mathsf{D}^{-1}[\mathbf{x}(t)])_{11} -v_y(t) \Delta t (\mathsf{D}^{-1}[\mathbf{x}(t)])_{12}
   + 2(\mathsf{g}^{-1}[\mathbf{x}(t)])_{11}\mathcal{N}_1(0,\Delta t) + 2(\mathsf{g}^{-1}[\mathbf{x}(t)])_{12} \mathcal{N}_2(0,\Delta t) \big]$;\\
   $v_y(t+\Delta t) = v_y(t) + \frac{1}{m}\big[ -v_x(t) \Delta t (\mathsf{D}^{-1}[\mathbf{x}(t)])_{21} -v_y(t) \Delta t (\mathsf{D}^{-1}[\mathbf{x}(t)])_{22}
   + 2(\mathsf{g}^{-1}[\mathbf{x}(t)])_{21}\mathcal{N}_1(0,\Delta t) + 2(\mathsf{g}^{-1}[\mathbf{x}(t)])_{22}\mathcal{N}_2(0,\Delta t) \big]$;
   }

   \Else{
   \If{$\alpha = 0$ }{
   $\bullet \ $ Friction term is constant, i.e.\\
   $v_x(t+\Delta t) = v_x(t) + \frac{1}{m}\big[ -v_x(t) \Delta t
   + g_{11}[\mathbf{x}(t)]\mathcal{N}_1(0,\Delta t) + g_{12}[\mathbf{x}(t)] \mathcal{N}_2(0,\Delta t) \big]$;\\
   $v_y(t+\Delta t) = v_y(t) + \frac{1}{m}\big[-v_y(t) \Delta t
   + g_{21}[\mathbf{x}(t)]\mathcal{N}_1(0,\Delta t) + g_{22}[\mathbf{x}(t)] \mathcal{N}_2(0,\Delta t) \big]$;
   }
   }

  $\bullet \ $ Increment position:\\
  $x(t + \Delta t) = x(t) + v_x(t+\Delta t)\Delta t$; \\
  $y(t + \Delta t) = y(t) + v_y(t+\Delta t)\Delta t$\;

  $\bullet \ $ Implement periodic boundary conditions:\\
  
  \If{$x(t+\Delta t) \geq L/2$ }{
   $\bullet \ $ $x(t+\Delta t) = x(t+\Delta t) - L$;
   }
   \If{$x(t+\Delta t) < L/2$ }{
   $\bullet \ $ $x(t+\Delta t) = x(t+\Delta t) + L$;
   }

   $\bullet \ $ Implement reflecting boundary condition:\\
   \If{$y(t+\Delta t) > L/2$ }{
   $\bullet \ $ $y(t+\Delta t) = L - y(t+\Delta t)$;
   }
   
   $\bullet \ $ Increment time: $t = t + \Delta t$\;
}

$\bullet \ $ Increment particle number: $n = n + 1$;
 
}
\end{algorithm}
\end{small}
\noindent where, for simplicity, we set $k_B T = 1$ in the case of $\alpha = 1$, leading to $\gamma = \mathsf{D}^{-1}$, and set $\gamma = k_B = 1$ in the case of $\alpha = 0$, leading to $\mathsf{D} = T$. As in Sec.~\ref{SM:SimulationScheme}, $t_f$ is the final time that the simulation is run to for each particle, $\Delta t$ is the simulation timestep, $N$ is the total number of particles, and $g_{ij}$ are the elements of the matrix $\mathsf{g} = \sqrt{2\mathsf{D}}$.

\subsection{A non-singular non-linear diffusion tensor}\label{SM:NonSingularDiffusionTensor}

The second case we consider is to directly regularise the diffusion tensor $\mathsf{D}'$ in Eq.~(13), which is singular at the cloak's inner boundary $r' = R_1$. This singularity resulted from the implicit blowing up of an infinitesimally small ``puncture'' in the original coordinates to a finite region (circular core of radius $R_1$) under the coordinate transformation, see Fig.~1 of the main text. This can be regularised from instead starting from a finite region of radius $\epsilon$ in the original coordinates \cite{kohn2008cloaking}, given by the following coordinate transformation,
\begin{equation}\label{eq:NonSingularTransformation}
    r' =
    \begin{cases}
        \frac{R_1}{\epsilon}r, & \quad r \leq \epsilon,\\
        \sqrt{\beta_\epsilon r^2 + \lambda_\epsilon}, & \quad \epsilon < r \leq R_2,\\
        r, & \quad r > R_2,
    \end{cases}
\end{equation}
where $\lambda_\epsilon = R_2^2 (R_1^2 - \epsilon_2)/(R_2^2 - \epsilon^2)$, $\beta_\epsilon = (R_2^2 - R_1^2)/(R_2^2 - \epsilon^2)$, and the azimuthal coordinate is once again invariant under the coordinate transformation, i.e.\ $\varphi' = \varphi$. The transformation in Eq.~(\ref{eq:NonSingularTransformation}) maps a region of finite radius $\epsilon$ to the inner core of radius $R_1$, i.e.\ $r' = R_1$ when $r = \epsilon$. In the limit $\epsilon \to 0$, one recovers the singular non-linear transformation of Eq.~(9), with the singular behaviour at $r = 0$ now more evident from Eq.~(\ref{eq:NonSingularTransformation}) when taking this limit.

\begin{figure}
    \centering
    \includegraphics[width=0.91\columnwidth, trim = 2cm 0cm 1.5cm 1cm, clip]{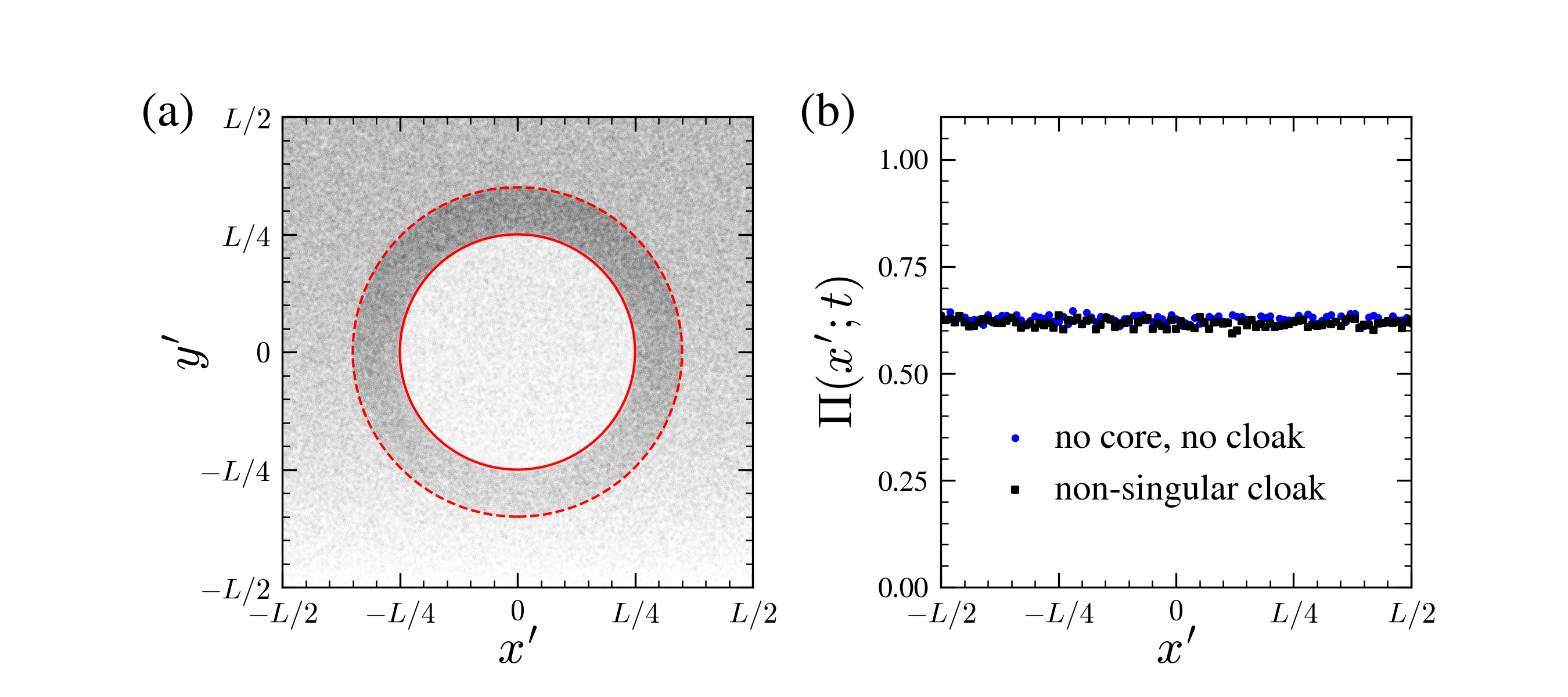}
    \caption{(a) Snapshot at time $t = 0.5$ from a simulation using the non-singular diffusion tensor, Eq.~(\ref{eq:TransformedDiffusionTensor_NonSingular}), derived in Sec.~\ref{SM:NonSingularDiffusionTensor}. Each data point in (a) represents the position of one of $N = 10^6$ particles. (b) Corresponding cumulative arrival distribution $\Pi(x;t)$, Eq.~(14), for the simulations of the non-singular diffusion tensor compared against that of no core. The simulation parameters used in both subfigures were $\Delta t = 10^{-5}$, $L=1$, $D_0 = 1$, $\epsilon = 0.2$, $R_1 = 0.25$, and $R_2 = 0.35$. Compared to the simulations of the singular diffusion tensor, Eq.~(13), illustrated in Fig.~2, there is significant particle penetration of the inner core. However, if the core has a homogeneous diffusivity of $D_0 R_1^2/\epsilon^2$, which is the case here, then the cumulative arrival distribution $\Pi(x;t)$ of the cloak better matches that of no core at all times, signifying that cloaking (in the sense that the particle density \textit{outside} the cloak remains invariant) is achieved to a greater degree.}
    \label{fig:NonSingularSnapshot}
\end{figure}

As in the main text, we calculate the Jacobian for this transformation in Cartesian coordinates. This results in
\begin{equation}\label{eq:NonSingularTransformation_Jacobian}
    \mathsf{J}
    = \frac{\partial (x,y)}{\partial (r,\varphi)}\frac{\partial (r,\varphi)}{\partial (r',\varphi')}\frac{\partial (r',\varphi')}{\partial (x',y')}
    = 
    \begin{cases}
        \frac{\epsilon}{R_1}\mathbbm{1}, & \quad r' \leq R_1,\\
        \mathsf{R}(\varphi') \mathrm{diag}\left( \frac{r'}{r \beta_\epsilon}, \frac{r}{r'}\right) \mathsf{R}(-\varphi'), & \quad R_1 < r' \leq R_2,\\
        \mathbbm{1}, & \quad r' > R_2,
    \end{cases}
\end{equation}
which has a piecewise-constant determinant given by
\begin{equation}\label{eq:NonSingularTransformation_detJ}
    \det{\mathsf{J}}
    = 
    \begin{cases}
        \frac{\epsilon^2}{R_1^2}, & \quad r' \leq R_1,\\
        \frac{1}{\beta_\epsilon}, & \quad R_1 < r' \leq R_2,\\
        1, & \quad r' > R_2.
    \end{cases}
\end{equation}
By bringing the general expression for the coordinate transformation of the Fokker-Planck equation into a form analogous to Eq.~(7) on a piecewise basis, as done for the singular case in the main text, one can identify the piecewise diffusion tensor for the non-singular transformation as
\begin{equation}\label{eq:TransformedDiffusionTensor_NonSingular}
    \mathsf{D}'_\epsilon
    = 
    D_0 \mathsf{J}^{-\mathrm{T}}\mathsf{J}^{-1}
    =
    \begin{cases}
        D_0 \frac{R_1^2}{\epsilon^2} \mathbbm{1}, & \quad r' \leq R_1,\\
        D_0 \beta_\epsilon \mathsf{R}(\varphi') \mathrm{diag}\left(\frac{r'^2 - R_1^2}{r'^2}, \frac{r'^2}{r'^2- R_1^2} \right) \mathsf{R}(-\varphi'), & \quad R_1 < r' \leq R_2,\\
        D_0 \mathbbm{1}, & \quad r' > R_2.
    \end{cases}
\end{equation}
This regularised diffusion tensor is more easily mimicked by a metamaterial in an experiment, since it removes the need for infinite diffusivity, but comes at the price of allowing particles to penetrate the protected region $r' \leq R_1$. However, if the core is of homogeneous diffusivity, then it is still possible to conceal it from an external observer. Specifically, \textit{given} a core of homogeneous diffusivity that we wish to conceal, cloaking is achieved by tuning the parameter $\epsilon$ in the cloak's spatially varying diffusivity such that the value $D_0 R_1^2 / \epsilon^2$ matches that of the core's given diffusivity.

Figure~\ref{fig:NonSingularSnapshot} shows a snapshot of the particle density and arrival distribution in the case of the non-singular diffusion tensor, Eq.~(\ref{eq:TransformedDiffusionTensor_NonSingular}), where particle positions evolve according to the overdamped Langevin equation (6) interpreted under the It\^{o} convention \cite{ito1944109}, $\alpha = 0$. The density is indistinguishable outside the cloak from that of a simulation of no core, as indicated by the arrival distribution more closely matching that of no core than for the singular diffusion tensor, Eq.~(13), presented in the main text, Fig.~2.

\subsection{Diffusion tensor for a spherical cloak}\label{SM:3D-DiffusionTensor}

Here, we discuss the feasibility of extending the results for the two-dimensional circular cloak of the main text to a three-dimensional spherical cloak. We start by considering an analogous transformation to the two-dimensional case, Eq.~(9), in the main text. In three dimensions, the radial coordinate once again transforms as
\begin{equation}\label{eq:NonLinearTransformation3D}
    r' =
    \begin{cases}
        \sqrt{\beta r^2 + R_1^2},& \quad r \leq R_2,\\
        r,& \quad r > R_2,
    \end{cases}
\end{equation}
where $\beta = (R_2^2 - R_1^2)/R_2^2$, and the azimuthal and polar coordinates are once again invariant, i.e. $\varphi' = \varphi$ and $\theta' = \theta$, respectively. In Cartesian coordinates $\mathbf{x}' = \left(x' = r'\cos(\varphi')\sin(\theta'), y' = r'\sin(\varphi')\sin(\theta'), z' = r'\cos(\theta') \right)$, the Jacobian for this three-dimensional non-linear transformation is
\begin{equation}\label{eq:NonLinearTransformation_Jacobian3D}
    \mathsf{J}
    = \frac{\partial (x,y,z)}{\partial (r,\varphi,\theta)}\frac{\partial (r,\varphi,\theta)}{\partial (r',\varphi',\theta')}\frac{\partial (r',\varphi',\theta')}{\partial (x',y',z')}.
\end{equation}
First, we have that    \begin{equation}\label{eq:Jacobian_xr_3D}
    \frac{\partial (x,y,z)}{\partial (r,\varphi,\theta)} = 
    \begin{pmatrix}
        \cos(\varphi) \sin(\theta) & -r\sin(\varphi)\sin(\theta) & r\cos(\varphi)\cos(\theta) \\
        \sin(\varphi)\sin(\theta) & r\cos(\varphi)\sin(\theta) & r\sin(\varphi)\cos(\theta)\\
        \cos(\theta) & 0 & -r\sin(\theta)
    \end{pmatrix}
    = 
    \mathsf{R}_z(\varphi)\mathsf{M}(\theta)\mathrm{diag}\left(1, r \sin(\theta), r \right),
\end{equation}
where
\begin{equation}
    \mathsf{M}(\theta) =
    \begin{pmatrix}
        \sin(\theta) & 0 & \cos(\theta) \\
        0 & 1 & 0\\
        \cos(\theta) & 0 & -\sin(\theta)
\end{pmatrix},
\end{equation}
and
\begin{equation}
    \mathsf{R}_z(\varphi) =
    \begin{pmatrix}
        \cos(\varphi) & -\sin(\varphi) & 0 \\
        \sin(\varphi) & \cos(\varphi) & 0 \\
        0 & 0 & 1
    \end{pmatrix}
\end{equation}
represents anticlockwise rotation in the $xy$-plane by $\varphi$. Then, combining the above with
\begin{equation}\label{eq:Jacobian_rr_3D}
    \frac{\partial (r',\varphi',\theta')}{\partial (r,\varphi,\theta)} = 
    \begin{cases}
        \mathrm{diag}\left(\frac{\beta r}{r'}, 1, 1\right), & R_1 < r' \leq R_2, \\
        \mathbbm{1}, & r' > R_2,
    \end{cases}
\end{equation}
we arrive at the following expression for the Jacobian of the three-dimensional transformation,
\begin{equation}\label{eq:Jacobian3Dfinal}
    \begin{split}
        \mathsf{J} &= 
        \begin{cases}
            \mathsf{R}_z(\varphi)\mathsf{M}(\theta)\mathrm{diag}\left(1, r \sin(\theta), r \right)\mathrm{diag}\left(\frac{r'}{\beta r}, 1, 1\right)\mathrm{diag}\left(1, \frac{1}{r' \sin(\theta')}, \frac{1}{r'} \right)\mathsf{M}^{-1}(\theta')\mathsf{R}_z^{-1}(\varphi'), & R_1 < r' \leq R_2, \\
            \mathsf{R}_z(\varphi)\mathsf{M}(\theta)\mathrm{diag}\left(1, r \sin(\theta), r \right)\mathbbm{1}\mathrm{diag}\left(1, \frac{1}{r' \sin(\theta')}, \frac{1}{r'} \right)\mathsf{M}^{-1}(\theta')\mathsf{R}_z^{-1}(\varphi'), & r' > R_2
        \end{cases}\\
        &= 
        \begin{cases}
            \mathsf{R}_z(\varphi')\mathsf{M}(\theta')\mathrm{diag}\left(\frac{r'}{\beta r}, \frac{r}{ r'}, \frac{r}{ r'} \right)\mathsf{M}(\theta')\mathsf{R}_z(-\varphi'), & R_1 < r' \leq R_2, \\
            \mathbbm{1}, & r' > R_2,
        \end{cases}
    \end{split}
\end{equation}
where we have also used that $r' = r$ for $r' \geq R_2$, as well as $\varphi' = \varphi$, $\theta' = \theta$, $\mathsf{M}^{-1}(\theta') = \mathsf{M}(\theta')$ and $\mathsf{R}_{z}^{-1}(\varphi') = \mathsf{R}_{z}(-\varphi')$.

Unlike that of the two-dimensional transformations considered in the main text and Sec.~\ref{SM:NonSingularDiffusionTensor}, the above Jacobian, Eq.~(\ref{eq:Jacobian3Dfinal}), is not piecewise homogeneous. For the three-dimensional spherical cloak, the determinant of the Jacobian is given by
\begin{equation}\label{eq:DetJacobian3D}
    \det{\mathsf{J}} = 
    \begin{cases}
        \frac{ r}{\beta r'}, & R_1 < r' \leq R_2, \\
        1, & r' > R_2.
    \end{cases}
\end{equation}
Thus, the generalisation to a three-dimensional spherical cloak on the basis of reading off the diffusion tensor from Eq.~(8) turns out not to be straightforward since, as for the two-dimensional case, this requires the determinant of the Jacobian to \emph{at least} be piecewise homogeneous. As for the two-dimensional case considered in the main text, a diffusion tensor of the form $\mathsf{D}' = D_0 \mathsf{J}^{-\mathrm{T}}\mathsf{J}^{-1}$ could be tested under the different integration conventions. A case could then be made \textit{a posteriori} if it has demonstrably good cloaking performance in a simulation, but this case would not be based on the same logic used to arrive at the diffusion tensor for the two-dimensional cloak. However, such a numerical investigation of the three-dimensional spherical cloak is beyond the scope of our work.


\end{document}